\documentclass[12pt,aps,nofootinbib,showpacs,groupedaddress,preprint]{revtex4}
\usepackage{epsfig}
\usepackage{dcolumn}   
\usepackage{bm}        
\usepackage{amssymb}

\begin{document}

\title{The Heavy Quark Potential as a Function of Shear Viscosity at
  Strong Coupling}

\preprint{BCCUNY-HEP/09-03}
\preprint{RBRC-780}

\author{Jorge Noronha}
\affiliation{Department of
Physics, Columbia University, 538 West 120$^{th}$ Street, New York,
NY 10027, USA}
\author{Adrian Dumitru}
\affiliation{Department of Natural Sciences, Baruch College, CUNY,
17 Lexington Avenue, New York, NY 10010, USA}
\affiliation{The Graduate School and University Center, City
  University of New York, 365 Fifth Avenue, New York, NY 10016, USA}
\affiliation{RIKEN BNL Research Center, Brookhaven National
  Laboratory, Upton, NY 11973, USA}

\begin{abstract}
We determine finite temperature corrections to the heavy-quark
(static) potential as a function of the shear viscosity to entropy
density ratio in a strongly coupled, large-$N_c$ conformal field
theory dual to five-dimensional Gauss-Bonnet gravity. We find that
these corrections are even smaller than those predicted by
perturbative QCD at distances relevant for small bound states in a
deconfined plasma. Obtaining the dominant temperature and viscosity
dependence of quarkonium binding energies will require a theory
where conformal invariance is broken in such a way that the free
energy associated with a single heavy quark is not just a pure
entropy contribution.
\end{abstract}

\date{\today}
\maketitle


\section{Introduction}

The Anti-de Sitter/Conformal Field Theory (AdS/CFT) correspondence
\cite{maldacena} relates correlation functions of local conformal
fields in 4-dimensional strongly-coupled non-Abelian plasmas to the
asymptotic behavior of fields defined in weakly-coupled, low energy
effective string theories in higher dimensions. The
conformal theories involved in the AdS/CFT correspondence depend
on the number of colors $N_c$ and on the t'Hooft coupling
$\lambda=g^2 N_c$. In particular, when $N_c\to \infty$ and
$g\to 0$ but $\lambda\gg 1$, the strongly coupled CFT in $D=4$
is dual to a weakly-coupled $D=10$ theory of (super)gravity. The
equivalence of strongly-coupled 4-dimensional $\mathcal{N}=4$
Supersymmetric Yang-Mills (SYM) to type IIB string theory on
AdS$_{5}\otimes S_{5}$ \cite{maldacena} has led to new insight
into the non-perturbative dynamics of strongly-coupled gauge theories at
finite temperature \cite{miscAdS}. For instance, it was shown that the
shear viscosity to entropy density ratio satisfies $4\pi\,\eta/s \geq
1$ in all gauge theories dual to supergravity \cite{viscobound}.

In general, due to the colossal number of possible vacua in the
current version of the string landscape \cite{Douglas:2006es}, one
may expect that higher derivative corrections to the gravity sector in
AdS$_{5}$ can occur. Using the relation $\sqrt{\lambda}=R^2/\alpha'$
(where $R$ is the radius of AdS$_5$), the $\mathcal{O}(\alpha')$
expansion in type IIB string theory becomes an expansion in powers
of $1/\sqrt\lambda$ in the SYM theory. Quartic corrections are known
to be present in closed superstring theory \cite{Grisaru:1986px}
(supersymmetry excludes terms corresponding to cubic powers of
Riemann tensors \cite{Hofman:2008ar}). In fact, it was shown in
\cite{Gubser:1998nz} that the leading corrections to the type IIB
tree level effective action are due to terms of the form
$\alpha\,'\,^{3} \,\mathcal{R}^4$, which in turn generate positive
corrections of $\mathcal{O}(\lambda^{-3/2})$ to $\eta/s$ that
preserve the viscosity bound \cite{Buchel:2004di}.

On the other hand, curvature squared interactions can be induced in
the effective 5-dimensional gravity sector by including the
world-volume action of D7-branes
\cite{Aharony:1999rz,Buchel:2008vz}, which are normally used in the
holographic description of the fundamental flavors in the dual gauge
theory \cite{Karch:2002sh}. It was shown in Refs.\
\cite{Brigante:2007nu,Kats:2007mq,brigantecausal} that 5-dimensional
gravity theories with curvature squared terms in the action are dual
to 4-dimensional superconformal theories where $\eta/s$ can be
lower than $1/(4\pi)$. Additionally, $\eta/s$ was found to be a very
simple analytical function of the new parameter associated with the
high derivative contributions (which is fully determined by the
central charges of the CFT). In fact, the very detailed study done
in Ref.\ \cite{Buchel:2008vz} confirmed (and extended) the initial
claim made in Ref.\ \cite{Kats:2007mq} that the viscosity bound
should be violated in superconformal theories with different central
charges.

In this paper, we use the gravity dual discussed in
\cite{Brigante:2007nu,brigantecausal}, which includes $\mathcal{R}^2$
corrections, to calculate the dependence of the heavy quark potential
at (moderately) short distances on $\eta/s$ in a strongly-coupled
non-Abelian plasma. The $Q\bar{Q}$ potential at finite temperature can
be calculated as a power series in $LT\ll 1$, where $L$ is the spatial
distance between the heavy quarks. It is shown that the potential
energy increases with $\eta/s$ and that the effective medium-induced
``screening'' of the attractive potential decreases much more rapidly with
increasing viscosity and quark mass at strong rather than at weak
t'Hooft coupling.

We would like to point out that we have limited our discussion to the
class of gravity theories that are dual to strongly coupled
superconformal gauge theories with non-equal central charges such as
those in \cite{Brigante:2007nu,brigantecausal,Kats:2007mq}. Other
corrections originating from $\mathcal{R}^4$ terms are not included in
our discussion. The combined effects of $\mathcal{R}^2$ and
$\mathcal{R}^4$ corrections to the heavy quark potential are left for
future work.

This paper is organized as follows. In the next section we review
how curvature squared corrections to the effective 5-dimensional
gravitational action affect the black brane horizon and,
consequently, lead to a modification of both thermodynamic and
transport properties of the dual D=4 CFT. In Section III we show how
these corrections affect the heavy quark potential at zero and
at finite temperature. Once the heavy quark potential is known,
in Section IV we determine the resulting binding energy of the
$Q\bar{Q}$ ground state and its
dependence on $\eta/s$. We close with a summary and outlook.

\section{$R^2$ corrections to the 5-dimensional gravitational action}

The effects of curvature squared corrections can be described by the
general action \cite{Brigante:2007nu,Kats:2007mq}
\begin{eqnarray}
 S &=& \frac{1}{16\pi G_5}\int d^{5}x \sqrt{-G}\left[\mathcal{R}
   +\frac{12}{R^2} \right. +\left. R^2 \left(c_1
   \mathcal{R}^2 +c_2 \mathcal{R}_{\mu\nu}\mathcal{R}^{\mu\nu} + c_3
   \mathcal{R}_{\mu\nu\lambda\rho}\mathcal{R}^{\mu\nu\lambda\rho}
   \right) \right]
\label{petrovkatsaction}
\end{eqnarray}
where $G_5=\pi R^3 /(2N_{c}^2)$ and $R$ is the radius of AdS$_5$
at leading order in $c_i$. The coefficients $c_{i}$ are expected to
be of $\mathcal{O}(\alpha')$, which means that
$\lim_{\lambda\to\infty} c_{i}=0$. However, at this order only $c_3$
is unambiguous because the coefficients $c_1$ and $c_2$ can be
arbitrarily modified via a simple redefinition of the metric
\cite{Brigante:2007nu,Kats:2007mq,Buchel:2008vz}.

The shear viscosity-to-entropy ratio, to first-order in $c_i$, was
found to be \cite{Brigante:2007nu,Kats:2007mq}
\begin{equation}
\frac{\eta}{s}=\frac{1}{4\pi}\left(1-8 c_3\right)+\mathcal{O}(c_i^2).
\label{petrovkatsetaovers}
\end{equation}
Therefore, the viscosity bound is violated when $c_3>0$. For
4-dimensional CFTs with AdS$_5$ gravity duals in the limit where
$\lambda \gg 1 $ and $N_c \to \infty$, one has
$c_3=\left(c-a\right)/(8c)+\mathcal{O}(1/N_c^2)$, where $a$ and $c$
are the central charges of the CFT \cite{Blau:1999vz}. For
$\mathcal{N}=4$ $SU(N_c)$ SYM $c=a$ exactly and the bound is
preserved, although there are superconformal theories in which
$\eta/s$ receives a correction of $\mathcal{O}(1/N_c)$ that violates
the bound \cite{Kats:2007mq,Buchel:2008vz}.

Gauss-Bonnet (GB) gravity \cite{Zwiebach:1985uq} is a special case
of the general action in (\ref{petrovkatsaction}) where $c_2=-4c_1$
and $c_1=c_3= \lambda_{GB}/2$, which gives the action
\begin{eqnarray}
 S_{GB} &=& \frac{1}{16\pi G_5}\int d^{5}x \sqrt{-G}\left[\mathcal{R}
   +\frac{12}{R^2}
 +\frac{\lambda_{GB}}{2}R^2\left(\mathcal{R}^2 -4
   \mathcal{R}_{\mu\nu}\mathcal{R}^{\mu\nu}  +
   \mathcal{R}_{\mu\nu\lambda\rho}\mathcal{R}^{\mu\nu\lambda\rho}
   \right)      \right] .
\label{GBaction}
\end{eqnarray}
For this particular combination of coefficients the metric
fluctuations around a given background have the same quadratic
kinetic terms as Einstein gravity (higher derivative terms cancel
\cite{Zwiebach:1985uq}). Another interesting feature of GB gravity
is that an exact black brane solution \cite{Cai:2001dz} is known for
$\lambda_{GB} \in (-\infty,1/4)$
\begin{equation}
\ ds^2 = -a^2 f_{GB}(U)dt^2 +\frac{U^2}{R^2}\,d\vec{x}^{\,2} +
\frac{dU^2}{f_{GB}(U)}~, \label{gaussbonnetmetric}
\end{equation}
where $a^2=\frac{1}{2}\left(1+\sqrt{1-4\lambda_{GB}}\right)$ and
\begin{equation}
\ f_{GB}(U)=\frac{U^2}{R^2}\frac{1}{2\lambda_{GB}} \left[ 1 -
  \sqrt{1-4\lambda_{GB} \left(1-\frac{U_h^4}{U^4}\right)}\right]~.
\label{gaussbonnetf}
\end{equation}
The parameter $a$ has the form above to make sure that the speed of
light at the boundary ($U\to \infty$) is unity. The horizon of the
GB black brane is the simple root of $f_{GB}$ located at $U=U_h$.
The plasma temperature in this case is
\begin{equation}
\ T=a \frac{U_h}{\pi R^2}
\label{tempGB}
\end{equation}
whereas the entropy density is
\begin{equation}
\ s=\frac{1}{4G_5}\left(\frac{U_h}{R}\right)^3=\frac{N_c^2 \pi^2
T^3}{2\,a^3} .\label{entropyGB}
\end{equation}

At this point the only formal constraint on the Gauss-Bonnet
coupling is that $\lambda_{GB}\in (-\infty,1/4)$. However, it was
shown in \cite{Brigante:2007nu} that
\begin{equation}
\frac{\eta}{s}=\frac{1}{4\pi}\left(1-4\lambda_{GB}\right),
\label{gauss-bonnetetaovers}
\end{equation}
to all orders in $\lambda_{GB}$. However, $\lambda_{GB}\leq 9/100$
or, equivalently, $4\pi\,\eta/s \geq 16/25$ in order to avoid
causality violation in the boundary \cite{brigantecausal}. In any
case, as was mentioned above, one should expect that
$|\lambda_{GB}| \sim \alpha'/R^2 \ll 1$ at strong t' Hooft coupling.
In this paper we take $\lambda_{GB}$ to be a free parameter
which parameterizes the ratio of shear viscosity to entropy density.

Note that the AdS radius in the GB geometry is not just $R$ but $aR$
\cite{Brigante:2007nu}. Thus, here we assume that the effective t'
Hooft coupling of the 4d CFT dual to the GB theory in Eq.\
(\ref{GBaction}) is $\lambda = R^4 a^4/\alpha'^2$. Moreover, the t'
Hooft coupling is assumed to be large such that qualitatively
meaningful results can be obtained at leading order in $\lambda$,
but finite~\cite{Gubser:2006qh}. The heavy-quark potential in the
strongly-coupled CFT only permits non-relativistic bound states, and
indeed bound states where the quarks are not localized over
distances smaller than their Compton wavelength, if the t' Hooft
coupling is not too large; c.f.\ Section~\ref{sec:IV}.

\section{$R^2$ corrections to the heavy quark potential}

We will be interested in the Wilson loop operator\footnote{Even
though the string dynamics can be in principle fully 10-dimensional,
here we consider only the dynamics corresponding to the 5
non-compact coordinates.}
\begin{equation}
\ W(C)=\frac{1}{N_c}{\rm Tr} \,P \,e^{i\int A_{\mu}dx^{\mu}}
\label{wilsonloop}
\end{equation}
where $C$ denotes a closed loop in spacetime and the trace is over the
fundamental representation of $SU(N_c)$. We
consider a rectangular loop with one
direction along the time coordinate $t$ and spatial extension
$L$. In the asymptotic limit $t\to \infty$, the vacuum
expectation value of the loop defines a static
potential via
\begin{equation}
\ \langle W(C) \rangle \sim e^{-t \,V_{Q\bar{Q}}(L)}~.
\label{qqbar1}
\end{equation}
Using somewhat loose language we call this the ``heavy-quark
potential''.

The expectation value of $W(C)$ can be calculated in the strongly
coupled $\mathcal{N}=4$ SYM theory using supergravity
\cite{Maldacena:1998im,Rey:1998ik}. According to the AdS/CFT
correspondence, an infinitely massive heavy quark in the fundamental
representation of $SU(N_c)$ in the $\mathcal{N}=4$ SYM theory is
dual to a classical string in the bulk that hangs down from a probe
brane at the boundary of AdS$_5$
\cite{Maldacena:1998im,Rey:1998ik} when $N_c\to \infty$ and $\lambda
\gg 1$ (supergravity approximation). Within this approximation, the
dynamics of the string (in Euclidean space) is given by the classical
Nambu-Goto action
\begin{equation}
\ S_{NG}=\frac{1}{2\pi \alpha'}\int d^2\sigma \sqrt{{\rm det} \,h_{ab}}
\label{nambugotoaction}
\end{equation}
where $h_{ab}=G_{\mu\nu}\partial_{a}X^{\mu}\partial_{b}X^{\nu}$
($a,b=1,2$), $G_{\mu\nu}$ is the background bulk metric,
$\sigma^{a}=(\tau,\sigma)$ are the internal world sheet coordinates,
and $X^{\mu}=X^{\mu}(\tau,\sigma)$ is the embedding of the string in
the 10-dimensional spacetime. In the supergravity approximation, since
the endpoint
of the string at the boundary carries fundamental charge, it is
natural to assume that
\begin{equation}
\ \langle W(C) \rangle \sim e^{-\Delta S_{NG}},
\end{equation}
where the loop $C$ is defined at the boundary of AdS$_5$. In the
equation above $\Delta S_{NG}$ is the regularized action, which
comes about after subtracting the infinite self-energy associated with two independent and infinitely massive quarks (two
straight lines that extend from $U=0$ to $U\to \infty$). Note that
this is consistent with the ideas behind holographic renormalization
\cite{holoanomaly}. The configuration that minimizes the action is a
curve that connects the string endpoints at the boundary and has a
minimum at some $U_*$ in AdS$_5$ \cite{Maldacena:1998im,Rey:1998ik}.

The potential for $\mathcal{N}=4$ SYM has the following simple analytical form
\cite{Maldacena:1998im}
\begin{equation}
\ V_{Q\bar{Q}}(L)=\frac{\Delta S_{NG}}{t}= -\frac{4\pi^2
  \sqrt{\lambda}}{\Gamma(1/4)^4} \frac{1}{L}~.
\label{maldacenaprop}
\end{equation}
The $\sim 1/L$ dependence is due to the conformal invariance of the
theory. Also, the potential is non-analytic in $\lambda$ while the
standard short-distance potential in perturbative QCD (pQCD) is, of
course, to leading order proportional to the coupling
\begin{equation}
\ V_{Q\bar{Q}}(L)\Big |_{QCD}=-\frac{g_{QCD}^2 C_F}{4\pi L} \simeq
-\frac{\lambda_{QCD}}{8\pi L}~,   \label{pQCDpot}
\end{equation}
where the latter form applies at large $N_c$ and $\lambda_{QCD}=g_{QCD}^2 N_c$.

One can generalize the calculations performed in
Refs.~\cite{Maldacena:1998im,Rey:1998ik} to include the effects from
curvature squared corrections given by, for instance, the
Gauss-Bonnet theory in Eq.~(\ref{GBaction}). The equations at zero
and at finite temperature are very similar and, thus, here we will
derive the general form of the equations and only later work out the
necessary details for each case.

In general, we have
\begin{equation}
\ {\rm det}\,h_{ab}=X^{'\,2}\cdot \dot{X}^2-(\dot{X}\cdot X')^2
\label{detstring}
\end{equation}
where $X^{'\,\mu}(\tau,\sigma)=\partial_{\sigma}X^{\mu}(\tau,\sigma)$
and $\dot{X}^{\mu} (\tau,\sigma) = \partial_{\tau} X^{\mu}
(\tau,\sigma)$. We choose a gauge where the coordinates of our static
string are $X^{\mu}=(t,x,0,0,U(x))$, where $\tau=t$ and
$\sigma = x$. Note that we use the Euclidean version of
eq.~(\ref{gaussbonnetmetric}) and, thus, at finite temperature the fields are
periodic in time with a period equal to $1/T$. In this case, one finds
\begin{equation}
\ S_{NG}=a\frac{t}{2\pi \alpha'}\int dx \sqrt{f_{GB}(U(x)) \frac{U^2
    (x)} {R^2}+U^{'\,2}(x)}~.
\label{generalNG}
\end{equation}
Note the presence of the prefactor $a(\lambda_{GB})$ in the equation
above. The Hamiltonian density associated with this action is
\begin{equation}
\ H_{NG}(x) = -\frac{U^2}{R^2} \frac{f_{GB}(U)} {\sqrt{f_{GB}(U)
    \frac{U^2} {R^2}+U^{'\,2}}}~,
\label{generalNG2}
\end{equation}
which is invariant under translations in $x$. In what follows we
denote the minimum of the U-shaped string at $x_*=0$ as $U_*$. One
can then compute $H_{NG}(x_*)$
\begin{equation}
\ H_{NG}(x_*)=-\sqrt{f_{GB}(U_*)\frac{U_*^2}{R^2}}~,
\label{generalNG1}
\end{equation}
which due to the translational symmetry is equal
to $H_{NG}(x)$ at any $x$. This allows us to solve for $x=x(U)$:
\begin{eqnarray}
 x(U) &=& \frac{R^2}{U_*}
\left[2\lambda_{GB}\left(1-\sqrt{1-4\lambda_{GB}\,\varepsilon} \right)
  \right]^{1/2}\, \int_{1}^{U/U_*}dy\,
  \left\{ \left[y^4 - y^2
    \sqrt{y^4-4\lambda_{GB}\left(y^4-1+\varepsilon\right)}\right]^2
  \right. \nonumber\\
&-& \left. \left[y^4-y^2 \sqrt{y^4 - 4\lambda_{GB}
      \left(y^4-1+\varepsilon\right)} \right]
\left[1-\sqrt{1-4\lambda_{GB}\,\varepsilon}\right]   \right\}^{-1/2}  ,
\label{profile1}
\end{eqnarray}
where $y_*\equiv U_h/U_*$ and $\varepsilon\equiv 1-y_{*}^4$. One of
the string endpoints is located at $x=-L/2$ while the other one is at
$x=L/2$. Thus, $U_*$ is related to $L$ via
\begin{eqnarray}
 \frac{L}{2} &=& \frac{R^2}{U_*} \left[2\lambda_{GB}
   \left(1-\sqrt{1-4\lambda_{GB}\,\varepsilon} \right) \right]^{1/2}
\,\int_{1}^{\infty}dy\,
  \left\{ \left[y^4 - y^2
    \sqrt{y^4-4\lambda_{GB}\left(y^4-1+\varepsilon\right)}\right]^2
  \right. \nonumber\\
&-& \left. \left[y^4-y^2 \sqrt{y^4 - 4\lambda_{GB}
      \left(y^4-1+\varepsilon\right)}\right]
\left[1-\sqrt{1-4\lambda_{GB}\,\varepsilon}\right]   \right\}^{-1/2}  ~.
\label{profile2}
\end{eqnarray}

Moreover, one can show that the regularized action is given by
\begin{eqnarray}
\frac{1}{2} \Delta S_{NG} &=& \frac{t
   \,a}{2\pi\,\alpha'}\int_{U_*}^{\infty}dU\left[1+\frac{1}
   {\frac{f_{GB}(U)U^2} {f_{GB}(U_*)U_{*}^2}-1}
   \right]^{1/2}-\frac{t \,a}{2\pi\,\alpha'}\int^{\infty}dU \label{regAction}\\
 &=& \frac{t \,a}{2\pi\,\alpha'}\,U_* \int_{1}^{\infty}dy \left\{
 \left[1+\frac{1}{\frac{f_{GB}(y)y^2}{f_{GB}(1)}-1}\right]^{1/2}
 -1 \right\}-\frac{t \,a}{2\pi\,\alpha'}U_*~, \label{NG2}
\end{eqnarray}
where
$y=U/U_*$ and
\begin{equation}
\ f_{GB}(y) = \frac{y^2}{R^2} \frac{U_{h}^2}{2\lambda_{GB}}
\left[1-\sqrt{1-4\lambda_{GB} \left(1-\frac{y_{h}^4}{y^4}\right)}\right]~.
\label{gaussbonnetfnew}
\end{equation}
We have regularized the action~(\ref{regAction}) by subtracting the
contribution of a straight string hanging down from the boundary
(corresponding to the infinite mass of the source). This also
subtracts a finite part of the action as determined by the lower limit
of the second integral from Eq.~(\ref{regAction}). We choose to
subtract (twice) the action at $T=0$, corresponding to
a straight string from $U=\infty$ to $U=0$. The free energy of the
$Q\bar{Q}$ pair is therefore identified with the entire
temperature-dependent contribution to the action.

At finite temperature, the free energy due to the heavy quarks (in a
color-singlet state) is given by the three-dimensional action of the
Wilson loop,
\begin{equation}
\ F_{Q\bar{Q}}=T \Delta S_{NG}~.
\label{FreeEnergyT}
\end{equation}
$F_{Q\bar{Q}}$ should not be interpreted as the heavy-quark
potential at finite temperature because it also contains an entropy
contribution~\cite{Shuryak:2004tx,Mocsy:2005qw} (especially at large
separation $L\to\infty$ where $F_\infty$ coincides with twice the
free energy due to a single heavy quark in the plasma; see
discussion below). We remove this entropy contribution at all $L$ by
defining
\begin{equation}
\ V_{Q\bar{Q}}=F_{Q\bar{Q}} - T\,\frac{\partial F_{Q\bar{Q}}}{\partial T} ~.
\label{potentialT}
\end{equation}
Thus, $V_{Q\bar{Q}}$ coincides with $F_{Q\bar{Q}}$ at short
distances (where temperature effects are absent) but approaches the
internal energy as $L\to\infty$~\cite{Mocsy:2005qw}.

\subsection{Heavy Quark Potential in the Vacuum}

The potential in the vacuum can be calculated to all orders in
$\lambda_{GB}$. In fact, when $T\to 0$ Eq.\ (\ref{profile2}) can be
easily solved for $U_*$
\begin{equation}
U_* = a(\lambda_{GB})\,\frac{2
  R^2}{L}\frac{\sqrt{2}\pi^{3/2}}{\Gamma(1/4)^2} ~.
\label{U*vacuum}
\end{equation}
This can be expressed as
\begin{equation}
\ U_* = a(\lambda_{GB}) \,\,U_*\Big |_{\rm Maldacena}
\label{U*vacuum1}
\end{equation}
where $U_*\Big |_{\rm Maldacena}$ is the result found in
\cite{Rey:1998ik,Maldacena:1998im}. Thus, we see that the bottom of
the U-shaped string approaches the boundary when $\lambda_{GB}$ goes
from $1/4$ to $-\infty$. The action for this configuration is
\begin{equation}
\Delta S_{NG} = -\frac{t}{L} \,\frac{4\pi^2 \sqrt{\lambda}}{\Gamma(1/4)^4}
\label{NGvacuumGB}
\end{equation}
where we used the previous definition $\sqrt{\lambda}=R^2
a^2/\alpha'$. Thus, the potential energy is
\begin{eqnarray}
V_{Q\bar{Q}}(L) &=& -\frac{1}{L} \, \frac{4\pi^2 \sqrt{\lambda}} {\Gamma(1/4)^4}
 \quad\quad\quad\quad\quad (\mbox{for}~{T=0}).
\label{eq:VacPot}
\end{eqnarray}
Both $\Delta S_{NG}$ and $V_{Q\bar{Q}}$ match the results of Ref.\
\cite{Maldacena:1998im} when expressed in terms of the appropriate
t' Hooft coupling in the gauge theory.

\subsection{Heavy Quark Potential at finite temperature}

We shall now proceed with the calculation of finite $T$ corrections
to the result above by expanding Eq.\ (\ref{profile2}) in powers of
$\delta=y_{*}^4$, assuming that $\delta\ll
(1/4-\lambda_{GB})/|\lambda_{GB}|$. This generalizes earlier results
for ${\cal N}=4$ SYM~\cite{Brandhuber:1998bs,Rey:1998bq,Albacete:2008dz}
to non-zero $\lambda_{GB}$. The boundary
condition~(\ref{profile2}) translates into
\begin{eqnarray}
LT =\frac{1}{2\sqrt{\pi}} \, \delta^{1/4} a^2\frac{\Gamma(3/4)}{\Gamma(5/4)}
\left[1-\frac{1}{5}\frac{\delta\,a^2}{\sqrt{1-4\lambda_{GB}}}\right]~.
\label{LT}
\end{eqnarray}
The limit $\delta\to0$ at fixed $\lambda_{GB}$ provides the leading
correction to the vacuum result from the previous section.
Expressing $\delta$ in terms of $LT$,
\begin{equation}
\delta = \frac{16\pi^2}{a^8} \, (LT)^4
\left(\frac{\Gamma(5/4)}{\Gamma(3/4)}\right)^4
\left[ 1 + \frac{64\pi^2}{5a^6} \frac{(LT)^4}{\sqrt{1-4\lambda_{GB}}}
\left(\frac{\Gamma(5/4)}{\Gamma(3/4)}\right)^4 \right]
\label{delta1_lowT}
\end{equation}
leads to
\begin{equation}
\ U_h= U_* \frac{2\sqrt\pi}{a^2} \, LT \,
\frac{\Gamma(5/4)}{\Gamma(3/4)}
\left[ 1 + \frac{16\pi^2}{5a^6} \frac{(LT)^4}{\sqrt{1-4\lambda_{GB}}}
\left(\frac{\Gamma(5/4)}{\Gamma(3/4)}\right)^4 \right] ~.
\label{U*new}
\end{equation}
The regularized action for this configuration is given by
\begin{eqnarray}
\Delta S_{NG} &=& -\frac{a\,U_*}{\pi \,\alpha'} \frac{1}{T}
 \frac{\sqrt{\pi}\,\,\Gamma(3/4)}{\Gamma(1/4)} \left[1+\frac{\delta\,
     a^2} {2\sqrt{1-4\lambda_{GB}}} +\mathcal{O}(\delta^2)\right] \\
&=& - \frac{2\sqrt{\lambda} \;}{LT} \left(
 \frac{\Gamma(3/4)}{\Gamma(1/4)}\right)^2
\left[ 1 + \frac{24\pi^2}{5} \frac{(LT)^4}{a^6 \sqrt{1-4\lambda_{GB}}}
\left( \frac{\Gamma(5/4)}{\Gamma(3/4)}\right)^4 \right]\nonumber\\
&& \quad\quad\quad\quad\quad\quad\quad~~~(\mbox{for}~
LT\to0)~.
\label{NG3}
\end{eqnarray}
In the last step we made use of Eqs.~(\ref{delta1_lowT})
and~(\ref{U*new}). At finite temperature we identify $\Delta S_{NG}$
with the free energy of the $Q\bar{Q}$ pair divided by the
temperature\footnote{Note that the entropy $S=-\partial
  F_{Q\bar{Q}}/\partial T$ for this configuration is indeed
  positive.}, and so Eq.~(\ref{potentialT}) leads to the following
potential:
\begin{eqnarray}
V_{Q\bar{Q}} &=& - \frac{2\sqrt{\lambda}\;}{L}
        \left( \frac{\Gamma(3/4)}{\Gamma(1/4)}\right)^2
        \left[ 1 - \frac{72\pi^2}{5} \frac{(LT)^4}{a^6 \sqrt{1-4\lambda_{GB}}}
\left( \frac{\Gamma(5/4)}{\Gamma(3/4)}\right)^4 \right]\nonumber\\
&& \quad\quad\quad\quad\quad\quad\quad~~~(\mbox{for}~
LT\to0)~.
\label{VQQ_lowT}
\end{eqnarray}
The first term coincides, of course, with the vacuum potential from
Eq.~(\ref{eq:VacPot}) while the second term corresponds to the
leading correction at small $LT$. Using Eq.\
(\ref{gauss-bonnetetaovers}), the potential can also be expressed in
terms of $\eta/s$
\begin{eqnarray}
V_{Q\bar{Q}} &=& - \frac{2\sqrt{\lambda}}{L}
        \left( \frac{\Gamma(3/4)}{\Gamma(1/4)}\right)^2
        \left[ 1 - \frac{576\pi^2}{5}
          \frac{(LT)^4}{\eta'}
          \frac{1}{\left( 1+\eta' \right)^3}
\left( \frac{\Gamma(5/4)}{\Gamma(3/4)}\right)^4 \right]~,
\label{VQQ_lowT_etas}
\end{eqnarray}
where $\eta' \equiv \sqrt{4\pi\frac{\eta}{s}}$. This expression
applies when the second term in the square brackets is a small
correction.

We observe that at fixed distance the potential decreases towards
higher temperature (however, its gradient increases in magnitude). We
compare to the behavior obtained from resummed pQCD where the
$Q\bar{Q}$ free energy at distances $m_DL \ll 1$ is given by
\begin{eqnarray}
F_{Q\bar{Q}} &=& - C_F \frac{g_{QCD}^2}{4\pi L} \left[ 1 -
\left(1-\frac{\xi}{6}\right) m_D \,L +
\frac{1}{2} \left(1-
       \frac{3\xi}{8}\right) m_D^2 \,L^2 + \cdots \right]~.
\label{FQQ_pQCD_etas}
\end{eqnarray}
This expression follows from the Fourier transform of the resummed
retarded propagator for static gluons~\cite{DGS,Dumitru:2009ni}.
Here, $m_D^2= g^2 N_c T^2/3 = \lambda_{QCD} T^2/3$ denotes the square of
the Debye screening mass at leading order. The parameter $\xi$ is
proportional to the product of $\eta/s$, expansion rate $\Gamma$,
and inverse temperature and is assumed to be small
\footnote{For very heavy quarks the time scale
  associated with the heavy quark bound state, $1/|E_{\rm bind}|$, is
  much shorter than the other time scales associated with temperature
  variations and the expansion rate. Thus, one can perform the
  calculations at fixed $T$ and set $\Gamma/T$ to be a constant.  The
  AdS/CFT result in Eq.\ (\ref{VQQ_lowT_etas}) should therefore be
  compared to the pQCD result assuming that $\xi$ is on the order of
  $\eta/s$ times a numerical coefficient.} \cite{xi_etas}:
\begin{equation}
\xi \sim \frac{\Gamma}{T}\, \frac{\eta}{s}~.
\end{equation}
If the entropy contribution is removed from eq.~(\ref{FQQ_pQCD_etas})
then medium induced screening effects are pushed to order $(m_DL)^2$
\cite{Dumitru:2009ni} and we obtain the following potential:
\begin{eqnarray}
V_{Q\bar{Q}} &=& - C_F \frac{g_{QCD}^2}{4\pi L} \left[ 1 -
\frac{1}{2} \left(1-\frac{3\xi}{8}\right) m_D^2 \,L^2 + \cdots \right]~.
\label{VQQ_pQCD_etas}
\end{eqnarray}
In qualitative agreement with~(\ref{VQQ_lowT_etas}), the potential
energy decreases (in magnitude) as $T$, and hence $m_D$, increases.
There is also qualitative agreement between
eqs.\ (\ref{VQQ_lowT_etas}) and~(\ref{VQQ_pQCD_etas}) in that the
``screening corrections'' (the second terms in the square brackets)
decrease as $\eta/s$ increases. However, note that the strong coupling
result in Eq.\ (\ref{VQQ_lowT_etas}) predicts a more rapid
disappearance of temperature effects as $LT\to0$ than the perturbative
QCD result shown in Eq.\ (\ref{VQQ_pQCD_etas}). The quartic dependence
on $LT$ in Eq.\ (\ref{VQQ_lowT_etas}) (also found in
Refs.\ \cite{Brandhuber:1998bs,Rey:1998bq}) originates from the
behavior of the metric near the horizon, i.e., the $(U_h/U)^4$ term in
Eq.\ (\ref{gaussbonnetf}) \footnote{In general, for black Dp-branes in
  asymptotically AdS$_{D}$ spaces (note that $D=p+2 \geq 5$) the
  correction would be $\sim (LT)^{p+1}$.}.

The free energy of a single heavy quark $F_Q$ in the plasma can also
be obtained from the regularized action in Eq.\ (\ref{NG2}). Due to
conformal invariance, it should be expected that $F_Q \sim T$ since
$T$ is the only energy scale available. In fact, one can simply take
the limit $U_* \to U_h$ in Eq.\ (\ref{NG2}) (straight string limit) to
show that
\begin{equation}  \label{FQ_CFT}
\ F_Q = - \frac{\sqrt\lambda}{1+\eta'}\, T~.
\end{equation}
Hence, $F_Q$ decreases in magnitude with increasing viscosity. This is
qualitatively similar to the behavior obtained from resummed
perturbation theory~\cite{Dumitru:2009ni} where
\begin{equation}  \label{FQ_pQCD}
\ F_Q = - \frac{1}{2} \alpha_s C_F m_D(T) \left(1-\frac{\xi}{6}+\cdots\right)
\end{equation}
at small $\xi$. Note that both (\ref{FQ_CFT}) and (\ref{FQ_pQCD})
are pure entropy contributions $\sim TS_Q = -\partial
F_Q/\partial\log T$ and so the potential energy of the quark in the
plasma vanishes once that is removed, according to
Eq.~(\ref{potentialT}). The $Q\bar{Q}$ potential at infinite
separation, $V_\infty$, is therefore zero.

Lattice data for the free energy of a static $Q\bar{Q}$ pair at
infinite separation, in SU(3) Yang-Mills theory as well as for 2, 2+1
and 3-flavor QCD~\cite{LattF}, can be parameterized
as~\cite{Mocsy:2005qw}
\begin{equation}  \label{FQ_Latt}
\ F_\infty(T) = 2F_Q(T) \simeq \frac{a}{T} - bT~,
\end{equation}
with $a\approx 0.08$~GeV$^2$ a constant of dimension two, not to be
confused with $a(\lambda_{GB})$ appearing in the
metric~(\ref{gaussbonnetf}), while $b$ is a dimensionless number.
The first term from Eq.~(\ref{FQ_Latt}) gives rise to a
non-vanishing $V_\infty(T)$ tied to the presence of an additional
dimensionful scale besides $T$.  In fact, for heavy quarks forming
very small bound states, the temperature dependence of the {\em
short-distance} potential is much smaller than that of
$V_\infty(T)$~\cite{Dumitru:2009ni} \footnote{The binding energy of
a quarkonium state is defined as the {\em eigenvalue} of the
  Hamiltonian relative to the potential at infinity (the latter
  corresponds to the sum of the potential energies of a $Q$ and a
  $\bar{Q}$ which do not interact with each other): $E_{\rm bind}=
  \langle\Psi\left|\hat{H}-V_\infty\right| \Psi\rangle -
  2m_Q$.}. Hence, we are presently unable to determine the dominant
temperature and viscosity dependent contribution to binding
energies, which would require a theory with broken conformal
invariance, perhaps along the lines of Ref.~\cite{Andreev:2006ct}
\footnote{In weakly coupled QCD, a contribution to the single-quark
  free energy of the form $F_Q\sim a/T$ could be generated by adding a
  non-perturbative contribution $m_G^2/(\vec{k}^2 + m_D^2)^2$ to the
  static gluon propagator; $m_G^2$ is a constant of dimension
  two~\cite{Megias:2007pq}. This also leads to a non-vanishing trace
  of the energy-momentum tensor~\cite{Megias:trace}.}. Rather, in the
following section we shall only compute the {\em eigenvalue} of the
Hamiltonian.

It is also interesting to recall that the expectation value of a
circular loop in ${\cal N}=4$ SYM at zero temperature is given
by~\cite{Drukker:2000rr}
\begin{equation}
\langle W\rangle_{\rm circ} = \exp\, \sqrt\lambda~,
\end{equation}
which agrees with our expression~(\ref{FQ_CFT}) if we identify the
expectation value of the loop with $\exp(-F_Q/T)$, where $1/T$ is
the length of the loop in the Euclidean time direction;
Eq.~(\ref{FQ_CFT}) also exhibits the dependence on the shear
viscosity in the large-$N_c$ limit and at sufficiently large t'
Hooft coupling $\lambda$.

\section{Heavy quark bound states} \label{sec:IV}

At small t' Hooft coupling bound states of heavy quarks
(``quarkonium'') have large Bohr radii $a_0 \gg 1/M_Q$ as compared
to the Compton wavelength of the quark and small binding energies
$|E_{\rm bind}|\ll M_Q$~\cite{Lucha:1991vn}. Therefore, a potential
model applies and the energy levels of the states can be obtained
from a Schr\"odinger equation. This is no longer the case if the
t'~Hooft coupling is very large. However, in practice one may take
$\lambda=g^2_{\rm YM} N_c\sim 5-10$~\cite{Gubser:2006qh} and the
numerical prefactor of the Coulomb-like $\sim 1/L$ potential
obtained via the AdS/CFT correspondence is smaller than unity.
Applying a potential model may therefore provide qualitatively
useful insight.

The heavy quark potential at $L\to 0$ is purely $\sim1/L$ for both
AdS/CFT and pQCD. As we saw in the previous section, the leading
corrections to the heavy quark potential in AdS/CFT and pQCD have
different powers of $LT$. We shall determine the energy levels for
both cases and check how they depend on $\eta/s$.

At short distances $V_{Q\bar{Q}}(r)=-A/r$, where
$A=4\pi^2 \sqrt{\lambda}/\Gamma(1/4)^4$ in GB and
$A=\lambda_{QCD}/\left(8\pi\right)$ for pQCD at large $N_c$. The
energy levels in the $\sim 1/r$ potential are
\begin{equation}
\ E_n^{T=0} = - M_Q \frac{A^2}{4n^2}~.
\label{energyvacuum}
\end{equation}
In what follows, we restrict ourselves to the $n=1$ ground state.
The ``Bohr radius'' of quarkonium is $a_0=2/(M_Q A) \ll 1/T$ at
sufficiently large quark mass. The wave function is
\begin{equation}
\psi_0=\frac{e^{-r/a_0}}{a_0^{3/2}\,\sqrt{\pi}}~.
\label{groundstatewave}
\end{equation}

At finite temperature the potentials have the form
\begin{equation}
\ V_{Q\bar{Q}}(r)= -\frac{A}{r}\left[1-B(rT)^\gamma\right]
\label{generalformpot}
\end{equation}
where for AdS/CFT $\gamma=4$ and
\begin{eqnarray}
\ B &=& \frac{72\pi^2}{5a^6 \sqrt{1-4\lambda_{GB}}}
\left( \frac{\Gamma(5/4)}{\Gamma(3/4)}\right)^4 \\
 &=& \frac{576\pi^2}{5} \frac{1}{\eta' (1+\eta')^3}
\left( \frac{\Gamma(5/4)}{\Gamma(3/4)}\right)^4~.
\end{eqnarray}
For pQCD $\gamma=2$ and
\begin{equation}
\ B= \frac{\lambda_{QCD}}{6}\left(1-\frac{3\xi}{8}\right)~.
\end{equation}
It is sufficient for our purposes here to compute the $T$-dependent
shift of the energy to leading order
\begin{equation}
\Delta E = \frac{4AB\,T^\gamma}{a_0^3}
\int_0^{\infty}dr\,r^{\gamma+1}\, e^{-2r/a_0} = AB
\frac{(a_0 T)^\gamma}{2^\gamma\, a_0} \, \Gamma(2+\gamma)
\label{energycorrection}
\end{equation}
and so the ground state energy level becomes
\begin{eqnarray}
\ E &=& - M_Q \frac{A^2}{4} \left( 1 - \frac{4}{2^\gamma} \frac{B}{A} \frac{(a_0
  T)^\gamma}{M_Q a_0} \,\Gamma(2+\gamma) \right)~.
\end{eqnarray}
Substituting for $A$, $B$ and $\gamma$ we obtain for GB
\begin{eqnarray}
\ E_{GB} &=& E_{GB}^{T=0} \left[1 - \frac{C}{\lambda^2}
\frac{1}{\eta'(1+\eta')^3} \frac{T^4}{M_Q^4} \right]
 ~,
\label{eq:E1AdS}
\end{eqnarray}
where
\begin{equation}
C = \frac{27}{256}\frac{\Gamma^{24}(1/4)}{\pi^{10}}  \approx 3
\times 10^7 ~,
\end{equation}
is a numerical constant and where $E^{T=0}$ denotes the ground state
energy in the ``Coulomb'' potential given in
Eq.~(\ref{energyvacuum}).

On the other hand, in pQCD
\begin{eqnarray}
\ E_{\rm pQCD} &=& - M_Q \frac{\lambda_{QCD}^2}{256 \pi^2} \left[1 -
\frac{128\pi^2}{\lambda_{QCD}}\left(1-\frac{3\xi}{8}\right)
\frac{T^2}{M_Q^2}
\right]\\
 &=& E_{\rm pQCD}^{T=0} \left[1 -
\frac{128\pi^2}{\lambda_{QCD}}\left(1-\frac{3\xi}{8}\right)
\frac{T^2}{M_Q^2} \right]  ~. \label{eq:E1pQCD}
\end{eqnarray}
As expected, the $T$-dependent shifts in
Eqs.~(\ref{eq:E1AdS},\ref{eq:E1pQCD}) exhibit a different dependence
on the t' Hooft coupling. However, the expression obtained from
AdS/CFT also drops more rapidly with $M_Q/T$ and with the
viscosity than predicted by pQCD.

\section{Summary and Outlook}
We have determined the dependence of the static $Q\bar{Q}$ potential
on the temperature $T$ and shear viscosity to entropy density ratio
$\eta/s$ in a conformal field theory dual to Gauss-Bonnet gravity on
AdS$_5$. We found that, with increasing viscosity, the screening of
the potential due to the thermal medium weakens and so the potential
energy increases in magnitude. Moreover, the free energy of a single
heavy quark decreases in magnitude with increasing viscosity. Both
observations are in qualitative agreement with expectations from
(``hard thermal loop'' resummed) perturbative QCD.

In fact, at short distances the medium-induced effects on quarkonium
binding energies are found to be very small, of order $\sim
(T/M_Q)^4 \times 1/\lambda^2 \eta'^4$, where
$\eta'\equiv\sqrt{4\pi\, \eta/s}$. The dominant temperature and
viscosity dependence of the binding energies therefore arises due to
the continuum threshold, i.e.\ from the value of the potential at
$L\to\infty$. Both pQCD (at leading order) as well as exactly
conformal gauge theories obtained using AdS/CFT, where $T$ is the
only dimensionfull scale available, can only generate a pure entropy
contribution to the free energy of the $Q\bar{Q}$ pair at infinite
separation, and so $V_\infty=0$ in both cases. It would be
interesting to construct a gravity dual for a field theory on the
boundary with a contribution of the form $\sim a/T$ to $F_\infty$ as
indicated by lattice QCD. This would provide a model for the
dominant $T$ and $\eta$ dependence of quarkonium binding energies in
a non-Abelian strongly-coupled plasma.

\section*{Acknowledgments}

J.N.\ acknowledges support from US-DOE Nuclear Science Grant No.\
DE-FG02-93ER40764. A.D.\ thanks R.~Pisarski and M.\ Strickland for
useful comments. J.N.\ thanks M.\ Gyulassy for interesting
discussions.

\end{document}